%% file: RVF.tex
\documentclass[12pt]{article}
\usepackage{amsmath}
\usepackage{cancel}
\usepackage{graphicx}

\newcommand{\bbms}{\bbs\ mixing}

\newcommand{\bbs}{\ensuremath{B_s\!-\!\Bbar{}_s\,}}

\newcommand{\Bbar}{\,\overline{\!B}}

\newcommand{\gev}{\, \mathrm{GeV}}
\newcommand{\tev}{\, \mathrm{TeV}}

\input econfmacros.tex
\def\Title#1{\begin{center} {\Large {\bf #1} } \end{center}}

\begin{document}

\Title{SUSY-breaking as the origin of flavour violation\footnote{Work done in collaboration with Lars Hofer, Ulrich Nierste and Dominik Scherer\cite{Crivellin:2008mq}}}

\bigskip\bigskip

%+\addtocontents{toc}{{\it D. Reggiano}}
%+\label{ReggianoStart}

\begin{raggedright}  

{\it Andreas Crivellin \index{Crivellin, A.}\\
Albert Einstein Center for Fundamental Physics\\
Institute for Theoretical Physics\\
CH-3012 Bern, Switzerland}
\bigskip\bigskip
\end{raggedright}

The smallness of the off-diagonal CKM elements and of the Yukawa couplings of the first two generations suggests the idea that these quantities are zero at tree-level and might be due to radiative corrections. In the MSSM self-energies involving the trilinear $A$-terms can lead to order one effects in the renormalization of the CKM elements and the light fermion masses. Thus, these quantities can be generated via loops ~\cite{Buchmuller:1982ye}. This radiative generation of the light fermion masses solves the SUSY CP problem~\cite{Borzumati:1999sp,Crivellin:2010ty} by a mandatory phase alignment between the $A$-terms and the effective Yukawa couplings.

However, the third generation fermion masses are too heavy to be loop generated and the successful bottom -- tau (top -- bottom) Yukawa coupling unification in SU(5) (SO(10)) GUTs suggests to keep tree-level third generation fermion masses. Thus we assume the following structure for the Yukawa couplings (and the CKM matrix) of the MSSM superpotential:
\begin{equation}
Y^{f}_{ij}  = Y^{f_{3}}\delta_{i3}\delta_{j3},\qquad V^{(0)}_{ij}  = \delta_{ij}\,.\label{eq:YukCKM}
\end{equation} 
This means that (in the language of \cite{D'Ambrosio:2002ex}) the global $[U(3)]^3$ flavor
symmetry of the gauge sector is broken down to $[U(2)]^3 \times [U(1)]^3$ by the Yukawa couplings of the third generation. Here the three $U(2)$ factors correspond to rotations of the
left-handed doublets and the right-handed singlets of the first two generation fermions in flavor space, respectively. 

The light quark masses and the off-diagonal CKM elements are generated by gluino-squark self-energies which means:
\begin{eqnarray}
\Sigma_{11}^{u,d\;LR} = m_{u,d},\qquad \Sigma_{22}^{u,d\;LR} = m_{c,s}, \qquad
   \frac{\Sigma_{13,23}^{d\;LR}}{m_{b}}-\frac{\Sigma_{13,23}^{u\;LR*}}{m_{t}} = V_{ub,cb}.
\label{CKMdown}
\end{eqnarray}
In the following we consider the limiting case in which the CKM matrix is solely generated by $\Sigma_{13,23}^{d\;LR}$ ($\Sigma_{13,23}^{u\;LR}$) which we call for obvious reasons CKM generation in the down(up)-sector. 

For CKM generation in the down-sector the most stringent constraint stems from an enhancement of $b\to s \gamma$ due to the off-diagonal element $\Delta_{23}^{d\;LR}$ in the squark mass matrix (left plot of Fig~1). Also $B_s\to\mu^+\mu^-$ receives sizable
contributions (even in the decoupling limit) from (loop-induced) flavor-changing
neutral Higgs couplings~\cite{Crivellin:2010er}. If in addition $\delta
_{23}^{d\;RL}\neq0$, \bbms\ is affected by double Higgs penguins as
well. In this way the \bbms\ phase which disagrees with the SM
expectation can be explained and interesting correlations with $B_s\to\mu^+\mu^-$ occur (see Fig.~2). 

In the case of CKM matrix generation in the up-sector,
the most stringent constraints stem from $\epsilon_K$ (right plot of Fig.~1) which receives additional contributions via a chargino box diagram involving the double mass insertion $\delta_{23}^{u\;LR}\delta_{13}^{u\;LR}$. At the same time the rare Kaon decays $K^+\to\pi^+\nu\overline{\nu}$ and $K_L\to\pi^0\nu\overline{\nu}$ receive sizable corrections 
which is very interesting for NA62.

In conclusion, since the off-diagonal CKM elements and the small quark masses can be
generated from loop diagrams while simultaneously obeying all FCNC
constraints (for SUSY masses at the TeV scale), the MSSM with RFV is a viable
alternative to the popular MFV variant of the MSSM. As opposed to the 
MFV-MSSM our RFV model is capable to explain the large \bbms\
phase favored by current data.

\begin{figure*}
\includegraphics[width=0.45\textwidth]{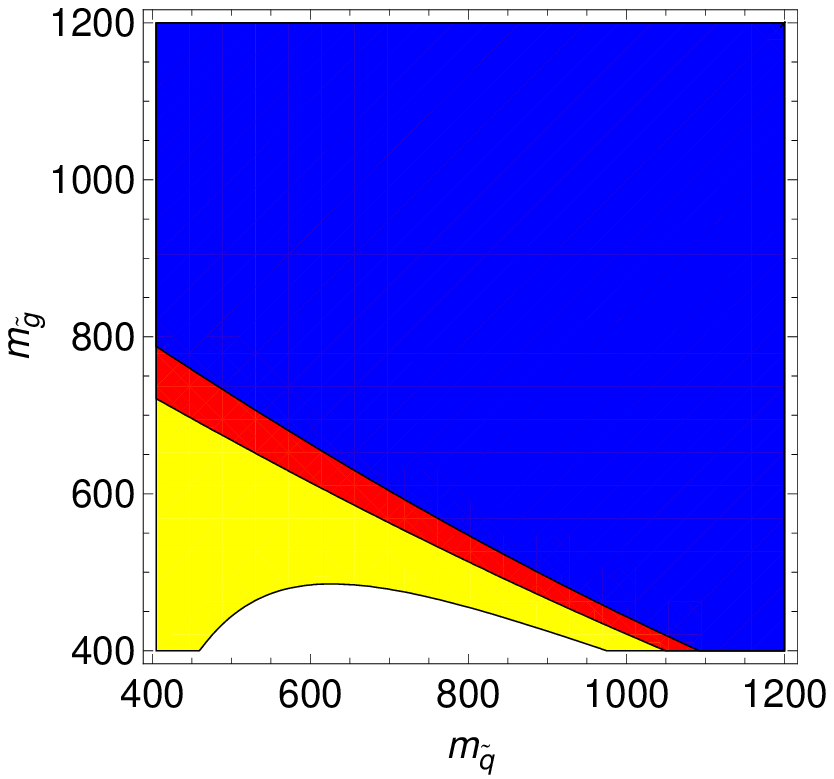}
\includegraphics[width=0.45\textwidth]{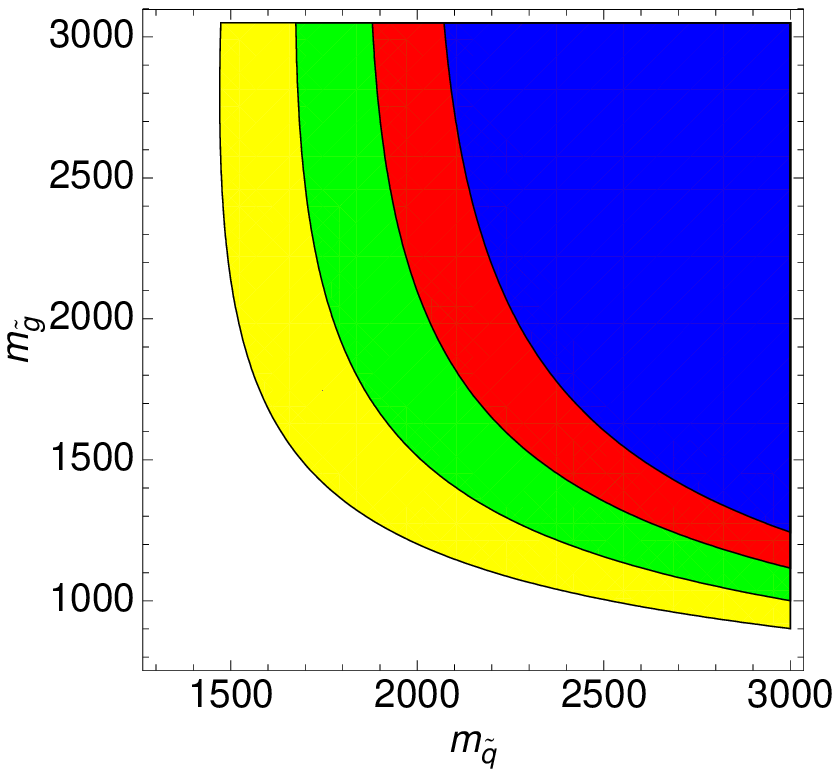}
\caption{\small{ Left: Allowed regions in the $m_{\tilde q}-m_{\tilde g}$
  plane. Constraints from $b\to s \gamma$ for different values of 
  $\mu \tan\beta$ assuming that the CKM matrix is generated
  in the down sector. We demand that the gluino contribution should not exceed the SM one. Yellow(lightest): allowed region for $\mu
  \tan\beta=30 \tev$, red: $\mu \tan\beta=0 \tev$ and blue(darkest):
  $\mu \tan\beta=-30 \tev$. Right: Allowed regions in the $m_{\tilde q}-m_{\tilde g}$ plane. Constraints from $K$--$\overline{K}$ mixing for different values of $M_2$ assuming that the CKM matrix is generated in the up sector. Yellow(lightest): $M_2=1000 \gev$, green: $M_2=750\gev$, red: $M_2=500\gev$ and blue(darkest): $M_2=250\gev$.}
\label{b-s-gamma-allowed}}
\end{figure*}
\begin{figure*}
\centering
\includegraphics[width=0.44\textwidth]{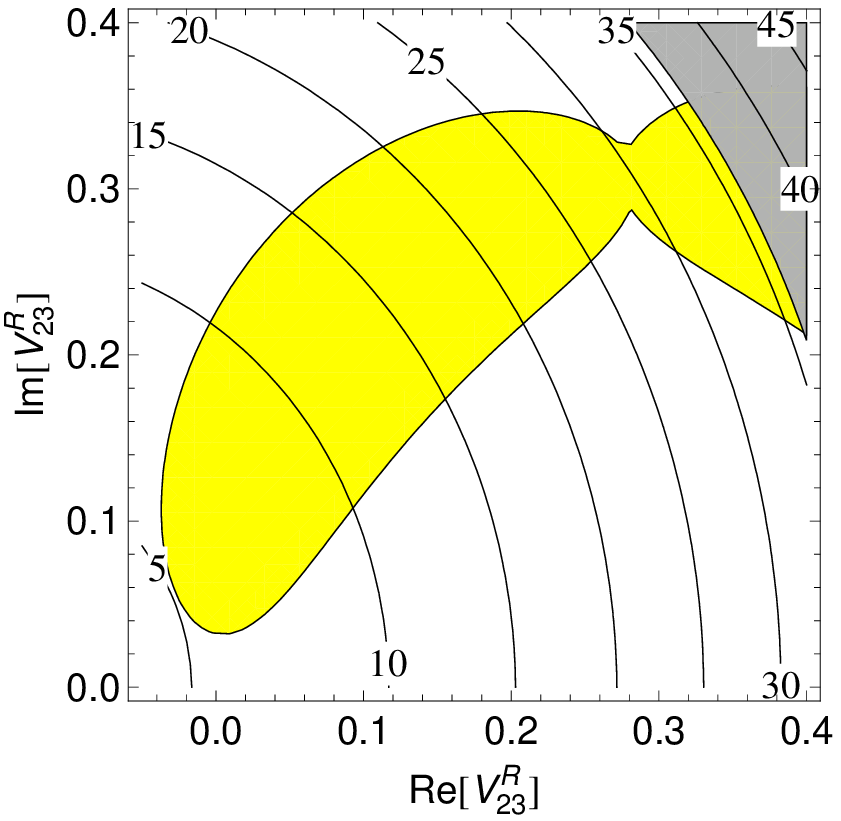}\hspace{0.08\textwidth}
\includegraphics[width=0.45\textwidth]{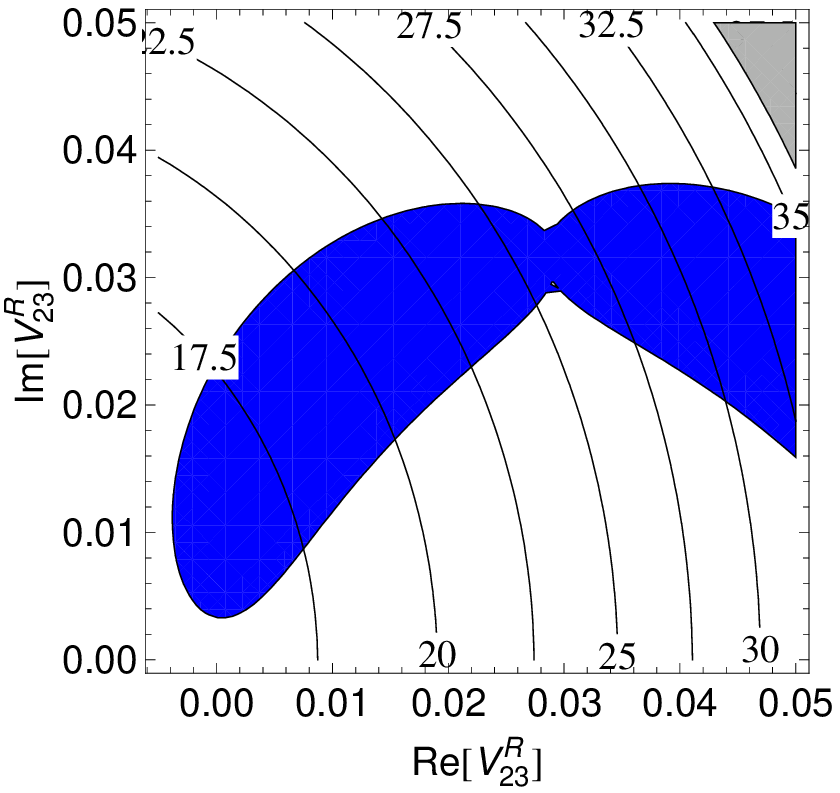}
\caption{Correlation between $B_s\to \mu^+\mu^-$ and \bbms\ for 
$\epsilon_b=0.0075$, $m_H=400\rm{GeV}$. 
Left plot, yellow: Allowed region from \bbms\ (95\% confidence
level) for $\tan\beta=11$. The contour-lines show $\rm{Br}[B_s\to
\mu^+\mu^-]\times10^9$. The grey area at the right side is excluded by the
Tevatron bound on $\rm{Br}[B_s\to\mu^+\mu^-]$. 
Right plot, blue: same for $\tan\beta=20$.\label{Bs-mixing2}}
\end{figure*}

\end{document}

%% file: econfmacros.tex
\def\beq{\begin{equation}}
\def\eeq#1{\label{#1}\end{equation}}
\def\eeqn{\end{equation}}

\def\beqa{\begin{eqnarray}}
\def\eeqa#1{\label{#1}\end{eqnarray}}
\def\eeqan{\end{eqnarray}}

\let\bar=\overbar

\def\Dslash{\not{\hbox{\kern-4pt $D$}}}
\def\dslash{\not{\hbox{\kern-2pt $\del$}}}

\def\msb{{\bar{\ssstyle M \kern -1pt S}}}